\newcommand{\vsini}{$v\sin i$}
\newcommand{\kms}{km s$^{-1}$}%
\newcommand{\teff}{$T_{\text{eff}}$}%
\newcommand{\bb}{$\langle B \rangle$}
\newcommand{\vmic}{$v_{\textrm{mic}}$}
\newcommand{\vmac}{$v_{\textrm{mac}}$}

\documentclass{aa}
\usepackage{graphicx}
\usepackage{color}
\usepackage{txfonts}
\begin{document}

  \title{Characterising the surface magnetic fields of T Tauri stars with high-resolution near-infrared spectroscopy}

\titlerunning{Magnetic fields of T Tauri stars from near-infrared spectroscopy}

  \author{
  A. Lavail\inst{1,2}
  \and
  O. Kochukhov\inst{1}
  \and
  G.A.J. Hussain\inst{2}
  }

  \institute{
   Department of Physics and Astronomy, Uppsala University, Box 516, SE-751 20 Uppsala, Sweden\\
   \email{alexis.lavail@physics.uu.se}
   \and
   European Southern Observatory, Karl-Schwarzschild-Strasse 2, 85748 Garching, Germany
  }

  \date{Received xx xx, 2019; accepted xx xx, 2019}

  \abstract%
{}
   {In this paper, we aim to characterise the surface magnetic fields of a sample of eight T Tauri stars from high-resolution near-infrared spectroscopy. Some stars in our sample are known to be magnetic from previous spectroscopic or spectropolarimetric studies. Our goals are firstly to apply Zeeman broadening modelling to T Tauri stars with high-resolution data, secondly to expand the sample of stars with measured surface magnetic field strengths, thirdly to investigate possible rotational or long-term magnetic variability by comparing spectral time series of given targets, and fourthly to compare the magnetic field modulus {\bb} tracing small-scale magnetic fields to those of large-scale magnetic fields derived by Stokes $V$ Zeeman Doppler Imaging (ZDI) studies.}
   {
   We modelled the Zeeman broadening of magnetically sensitive spectral lines in the near-infrared $K$-band from high-resolution spectra by using magnetic spectrum synthesis based on realistic model atmospheres and by using different descriptions of the surface magnetic field. We developped a Bayesian framework that selects the complexity of the magnetic field prescription based on the information contained in the data. }
   {We obtain individual magnetic field measurements for each star in our sample using four different models. We find that the Bayesian Model 4 performs best in the range of magnetic fields measured on the sample (from 1.5~kG to 4.4~kG). We do not detect a strong rotational variation of {\bb} with a mean peak-to-peak variation of 0.3~kG. Our confidence intervals are of the same order of magnitude, which suggests that the Zeeman broadening is produced by a small-scale magnetic field homogeneously distributed over stellar surfaces. A comparison of our results with mean large-scale magnetic field measurements from Stokes $V$ ZDI show different fractions of mean field strength being recovered, from 25--42\% for relatively simple poloidal axisymmetric field topologies to 2--11\% for more complex fields.}
   {}

   \maketitle
%
\section{Introduction}
With an ever improving understanding of stellar magnetism, we now realise that magnetic fields have a paramount impact throughout the entire life of a star. Particularly on the pre main-sequence (PMS), where stars are relatively cool and magnetic fields seem ubiquitous, magnetism has a significant impact on stars themselves, their formation, accretion properties, rotation rate, flares, and wind characteristics among others. It also influences potential orbiting exoplanets, especially their chemical evolution and habitability.

The origin of stellar magnetic fields and their evolution along the PMS, from stars that evolve from a fully-convective regime to a partially radiative and then eventually to a fully radiative regime, is not fully understood. While cool PMS T Tauri stars exhibit ubiquitous magnetic fields, it seems that most stars, that evolve into PMS Herbig Ae/Be stars and eventually to A/B type main-sequence stars, lose their fields at some point \citep{2013MNRAS.429.1001A,2019MNRAS.483.3127S}. The 5--10\% of stars remaining magnetic on the main sequence display simple, fossil fields that are not maintained by an active dynamo. Studying the evolution of magnetic fields in the T Tauri star regime in more detail can allow us to understand when and how fast this transition occurs, and perhaps identify different populations with distinct magnetic properties. This could improve our understanding of the mechanisms and timescales at play regarding the evolution and survival of the magnetic fields in the later evolutionary stages \citep{2019EAS....82..345A}.

There are two direct methods widely used to measure stellar magnetic fields that both rely on the Zeeman effect. The first is the measurement of Zeeman broadening, or splitting of spectral lines in intensity spectra, and the second is the analysis of the time series of polarised spectra using the Zeeman Doppler Imaging (ZDI) method. These two methods complement each other well. On the one hand, the first method provides a robust estimate of the magnetic field strength integrated over the stellar surface, and even takes small-scale magnetic structures into account. However, it provides almost no information on magnetic field topology. On the other hand, ZDI can reveal the topology of the large-scale magnetic field component. Yet, this method tends to miss some small-scale field structures due to the cancellation of opposite polarities, particularly when only circular polarisation is used. This leads to the severe underestimation of the surface field strength. The two methods have usually not been applied to the same datasets as spectropolarimeters mostly operate in the optical regime (up to 1050~nm), whereas the Zeeman broadening effect increases with wavelength and therefore is best studied at longer wavelengths (for example, in the near-infrared at $\approx$1500~nm or $\approx$2200~nm). In a few cases it was possible to obtain both types of measurements from the same data \citep{2017ApJ...835L...4K,2019MNRAS.tmp..410H,2019ApJ...873...69K} by capitalising on the wide wavelength coverage of optical spectropolarimeters, such as the ESPaDOnS (Echelle SpectroPolarimetric Device for the Observation of Stars) instrument \citep{donati2003-espadons} at the Canada--France--Hawaii Telescope. The possibilities for self-consistent magnetic analyses will be greatly improved with the advent of a new generation of near-infrared spectropolarimeters such as SPIRou (SpectroPolarim\`etre Infra-Rouge) at the Canada--France--Hawaii Telescope and CRIRES+ (the upgraded CRyogenic high-resolution InfraRed Echelle Spectrograph) at the European Southern Observatory (ESO) Very Large Telescope \citep{2014SPIE.9147E..15A,2016SPIE.9908E..0ID}.

A substantial sample of T Tauri stars has been studied using Zeeman broadening \citep{1992ApJ...390..622B,1999A&A...341..768G,2005ApJ...635..466Y,2007ApJ...664..975J,2008AJ....136.2286Y,2011ApJ...729...83Y,2017A&A...608A..77L,2018ApJ...853..120S}. A once smaller, but rapidly expanding sample of PMS stars has had its range of large-scale fields characterised with spectropolarimetry and mapped using ZDI \citep[][and references therein]{2019villebrun,2018MNRAS.480.1754N,2019MNRAS.tmp..410H}. It is worthwhile to note that all these ZDI studies only used Stokes $IV$ parameters.

In this work we aim to expand the sample of T Tauri stars with a magnetic field characterised from high-resolution near-infrared spectroscopy. Our sample includes stars that have been studied with the ZDI technique, allowing us to compare, in detail, the results obtained with the two methods, which are likely to be sensitive to magnetic fields at different spatial scales. Additionally, all but one star were observed repeatedly, enabling us to characterise rotational variability of the magnetic field responsible for Zeeman broadening. We also advance the methodology of magnetic measurements by testing different prescriptions of the distribution of magnetic field strengths over the stellar surface, including a data-driven approach which lets the information contained in the observed spectra determine the appropriate parameterisation of the magnetic field strength distribution.

This paper is structured as follows. The observations and the data reduction process are described in Sect.~\ref{section:observations}, Sect.~\ref{section:magnetic-spectrum-synthesis} explains the methods of our analysis, we present our results in Sect.~\ref{section:results} and finally discuss them in Sect.~\ref{section:discussion}

\section{Observations and data reduction}
\label{section:observations}
We acquired high-resolution spectra with the CRIRES spectrograph \citep{2004SPIE.5492.1218K} mounted at one of the Nasmyth foci of the 8.2 m Very Large Telescope Unit Telescope 1 at the ESO Paranal observatory. Our observations were obtained between mid-2008 and mid-2009, and were performed in nodding mode in order to improve the sky subtraction. The log of our observations is presented in Table~\ref{table:observations}. A 0.2$\arcsec$--wide slit yielded a resolving power of $ R \equiv \lambda / \Delta \lambda \approx 10^5$. The spectra were recorded on a mosaic of four Aladdin~III~InSb detectors amounting to an effective $4096 \times 512$ pixels focal plane detector array. The data were reduced with the standard ESO CRIRES pipeline. We then used the {\tt MOLECFIT} software \citep{2015A&A...576A..77S} to remove the telluric lines from the reduced spectra, and simultaneously improve the wavelength solution obtained by the CRIRES pipeline. {\tt MOLECFIT} retrieves the atmospheric and telescope conditions at the time and location of the observations, and fits the telluric lines in the observed spectra by performing a high-resolution radiative transfer modelling of the terrestrial atmosphere along the line of sight. An example of telluric removal is illustrated in Fig.~\ref{figure:molecfit}. Recent work by \citet{2019A&A...621A..79U} demonstrated that {\tt MOLECFIT} is an excellent tool for removing telluric lines from CRIRES spectra and that it performs better than the classical approach of using a telluric standard star. In our correction, each of the CRIRES detectors was treated independently. In the final reduction step, the spectra from each detector were normalised using a custom routine, which fitted continuum level with a third order polynomial.

\begin{figure*}
    \centering
    \includegraphics[width=2.0\columnwidth]{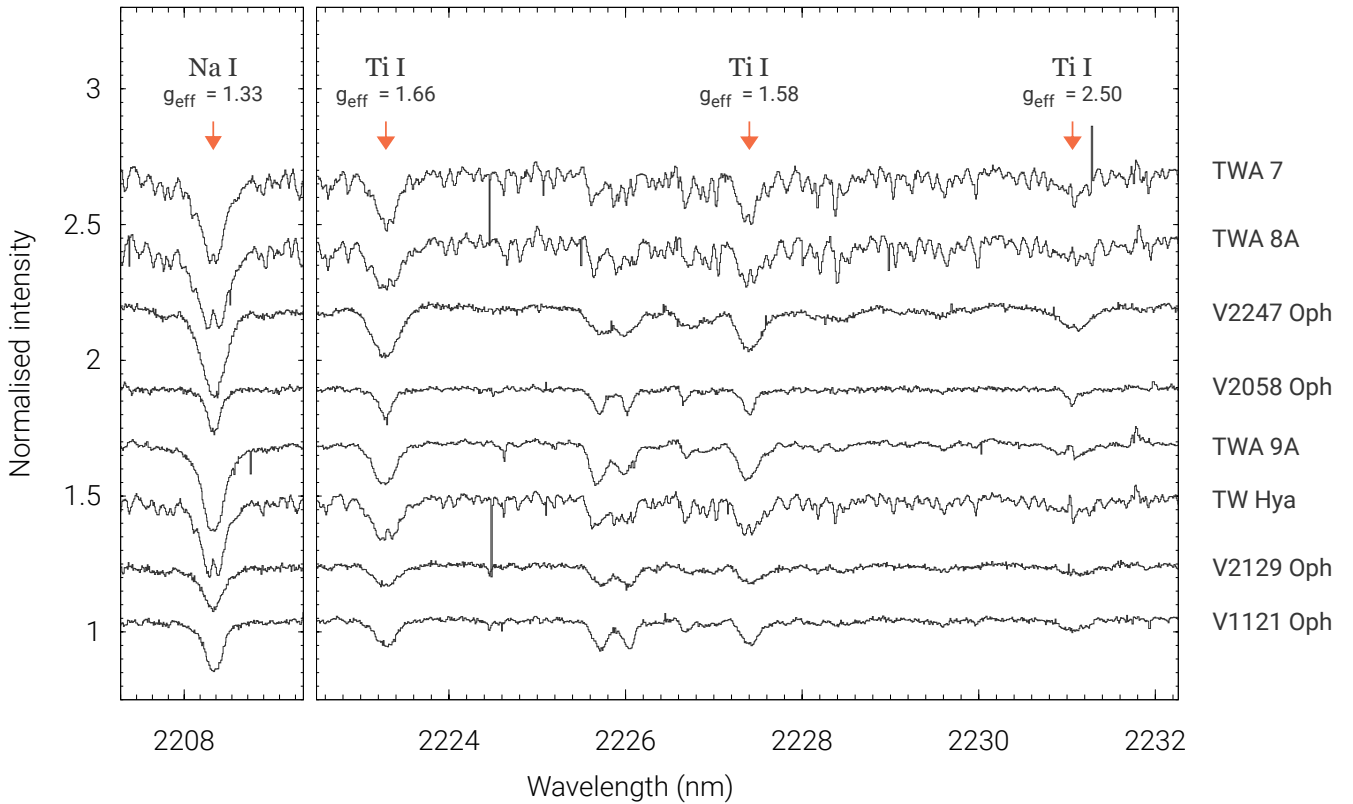}
    \caption{
        Highest S/N normalised spectra in wavelength setting $25/-1/i$ for each target in our sample. The left panel shows the strong \ion{Na}{I} line present on the first CRIRES detector. The right panel shows the entire second CRIRES detector covering several magnetically sensitive \ion{Ti}{I} lines. The spectral lines used in this analysis and their respective effective Land\'e factors $g_{\text{eff}}$ are indicated at the top.
    }
    \label{figure:spec}
    \end{figure*}
The stars in our sample were observed in either one or two CRIRES standard wavelength settings. All targets were observed in the first wavelength setting $25/-1/i$ at the reference wavelength 2242.8~nm. This setting contains the strong, magnetically sensitive \ion{Ti}{i} lines at $\lambda$ 2223.3, 2227.4, and 2231.1~nm, as well as the strong \ion{Na}{i} line at $\lambda$ 2208.4~nm. We used this set of lines to determine the mean magnetic field modulus {\bb}. All targets but TW~Hya were observed in the second wavelength setting $24/-1/n$ at the reference wavelength 2329.3~nm. This setting covers several lines from the CO band-head, which are magnetically insensitive and therefore are useful for constraining different sources of non-magnetic broadening such as {\vsini}. The spectra yielding the highest signal-to-noise ratio (S/N) of each star acquired in the wavelength setting $25/-1/i$ are displayed in Fig.~\ref{figure:spec}. In the particular case of the resolved binary TWA 9 AB, the slit was aligned along the two components, separated by 6\arcsec, and the spectrum of the primary TWA 9A was extracted individually.

Except for V2247~Oph, stars in our samples were observed repeatedly (2--5 times), in order to investigate potential rotational variability. Some observations were acquired during consecutive nights while others were months apart. The observing dates are indicated in Table~\ref{table:observations}.
%
\begin{table*}[!ht]
\caption{Journal of observation.}
\label{table:observations}
\centering
\begin{tabular}{l c c c c c c }     
\hline\hline
Target ID   & Other ID  & UT date       & UT time at start & Wavelength setting & Integration time & S/N \\
\hline
\object{TWA 7}       & CE Ant    & 2008-12-24    & 06:43:41  & 2242.8 nm & $3 \times 45$ s   & 61    \\
            &           & 2008-12-24    & 07:03:49  & 2242.8 nm & $3 \times 45$ s   & 63    \\
            &           & 2009-02-12    & 07:40:19  & 2242.8 nm & $3 \times 45$ s   & 71    \\   
            &           & 2008-12-24    & 07:25:53  & 2242.8 nm & $3 \times 45$ s   & 62    \\
            &           & 2009-02-24    & 02:15:43  & 2242.8 nm & $3 \times 45$ s   & 70    \\
            &           & 2008-12-24    & 07:46:48  & 2329.3 nm & $3 \times 45$ s   & 42    \\
\object{TWA 8A}      & V550 Hya  & 2009-02-26    & 03:48:54  & 2242.8 nm & $3 \times 90$ s   & 74    \\
            &           & 2009-02-27    & 02:07:57  & 2242.8 nm & $3 \times 90$ s   & 73    \\
            &           & 2009-02-28    & 02:02:54  & 2242.8 nm & $3 \times 90$ s   & 76    \\
            &           & 2009-02-24    & 02:54:02  & 2329.3 nm & $3 \times 90$ s   & 48    \\
\object{V2247 Oph}   & ROX 21    & 2008-08-24    & 00:33:47  & 2242.8 nm & $3 \times 180$ s  & 164   \\
            &           & 2009-04-20    & 05:13:07  & 2329.3 nm & $3 \times 180$ s  & 153   \\
\object{V2058 Oph}   & ROX 6     & 2008-07-09    & 01:04:52  & 2242.8 nm & $3 \times 90$ s   & 157   \\
            &           & 2008-07-10    & 00:17:30  & 2242.8 nm & $3 \times 90$ s   & 202   \\
            &           & 2008-08-26    & 00:32:50  & 2242.8 nm & $3 \times 90$ s   & 183   \\
            &           & 2008-08-29    & 01:03:14  & 2329.3 nm & $3 \times 90$ s   & 127   \\
\object{TWA 9A}      & V1239 Cen & 2008-07-08    & 00:55:51  & 2242.8 nm & $3 \times 180$ s  & 85    \\
            &           & 2009-02-27    & 02:56:48  & 2242.8 nm & $3 \times 180$ s  & 204   \\
            &           & 2009-02-28    & 02:36:36  & 2242.8 nm & $3 \times 180$ s  & 226   \\
            &           & 2009-02-24    & 03:23:08  & 2329.3 nm & $3 \times 180$ s  & 162   \\
\object{TW Hya}      & TWA 1     & 2008-12-24    & 08:10:07  & 2242.8 nm & $3 \times 90$ s   & 89    \\
            &           & 2009-02-12    & 07:05:58  & 2242.8 nm & $3 \times 90$ s   & 99    \\
            &           & 2009-02-15    & 06:24:19  & 2242.8 nm & $3 \times 90$ s   & 102   \\
\object{V2129 Oph}   & ROX 29    & 2008-08-24    & 00:01:29  & 2242.8 nm & $3 \times 60$ s   & 177   \\
            &           & 2008-08-26    & 00:01:58  & 2242.8 nm & $3 \times 60$ s   & 146   \\
            &           & 2008-08-28    & 23:40:46  & 2242.8 nm & $3 \times 60$ s   & 147   \\
            &           & 2008-08-29    & 00:06:20  & 2329.3 nm & $3 \times 60$ s   & 119   \\
\object{V1121 Oph}   & AS 209    & 2008-08-23    & 23:18:49  & 2242.8 nm & $3 \times 45$ s   & 128   \\
            &           & 2008-08-25    & 23:38:39  & 2242.8 nm & $3 \times 45$ s   & 201   \\
            &           & 2008-07-24    & 01:47:46  & 2242.8 nm & $3 \times 45$ s   & 184   \\
            &           & 2008-05-24    & 03:35:16  & 2329.3 nm & $3 \times 45$ s   & 140   \\
\hline
\end{tabular}
\end{table*}

\section{Magnetic spectrum synthesis}
\label{section:magnetic-spectrum-synthesis}
Our analysis relies on a direct comparison between observations and theoretical synthetic spectra. These spectra were calculated using the {\tt Synmast} code \citep{OK2007pms,2010A&A...524A...5K}, which numerically solves the polarised radiative transfer equation for a given stellar model atmosphere, spectral line list and magnetic field vector under the assumption of local thermodynamical equilibrium (LTE). {\tt Synmast} incorporates a modern molecular equilibrium solver \citep{2017A&A...597A..16P}, allowing to treat both hot and cool star atmospheric conditions. The plane-parallel hydrostatic 1-D stellar atmospheres employed in our study were taken from the {\tt MARCS} model atmospheres grid \citep{marcs2008}\footnote{\url{http://marcs.astro.uu.se/}}. The {\tt MARCS} atmospheres are not magnetic and hence might not be fully realistic models of magnetised T Tauri stars atmospheres where the single surface temperature assumption might be invalid. However, as we lack appropriate theoretical models (such as 3D magnetoconvection atmospheric simulations for classical T Tauri stars), it is extremely challenging to model the impact of the magnetic field in this context. Nevertheless, the impact of multiple temperatures on the stellar surface would be rather mild on our final results as our analysis is quite insensitive to changes in effective temperature (as detailed in Sect.~\ref{section:results}).

For the purpose of modelling the Zeeman broadening, we divided the visible stellar hemisphere in seven annular regions of equal area and computed the local intensity spectra for each region. These local spectra were then broadened, disk-integrated, and normalised to the continuum. Here we generated a grid of synthetic spectra for a purely radial, homogeneous magnetic field, ranging from 0 to 14~kG with 1~kG steps. These synthetic spectra were later linearly combined to simulate more complex surface magnetic field strength distributions, following different prescriptions discussed in Sect.~\ref{subsection:magnetic_field_analysis}. Calculations adopted solar abundances according to \citet{2009ARA&A..47..481A}.

The line list used in this study was extracted from the {\tt VALD3} database \citep{vald2015}\footnote{\url{http://vald.astro.uu.se}}. We corrected this list in the following two occurrences where data were missing or produced a bad fit to the observed spectra. One of the four spectral lines used to determine the magnetic field modulus, the \ion{Na}{I} line at $\lambda$ 2208.4~nm, has no van der Waals broadening coefficient in the {\tt VALD3} database. In this case {\tt Synmast} defaults to the van der Waals broadening treatment according to \citet{1992oasp.book.....G}, which appears to underestimate the broadening. We therefore adopted an extended broadening coefficient of $1300.210$ calculated by P. Barklem (priv. comm.) using the ABO theory \citep{1995MNRAS.276..859A,1997MNRAS.290..102B}. This changes significantly the shape of the \ion{Na}{I} line and produces substantially broader wings. Furthemore, we noticed that the synthetic spectra showed systematically stronger blending lines in the blue wing of the \ion{Ti}{I} line at $\lambda$ 2227.4~nm when compared to observations. This was due to two OH transitions around $\lambda$ 2227.2~nm. We reduced the oscillator strengths of these two transitions by 7 and 5 dex respectively to reach a good fit to observations.

\subsection{Stellar parameters}
\label{subsection:stellar-parameters}
\begin{figure*}
\centering
\includegraphics[width=1.6\columnwidth]{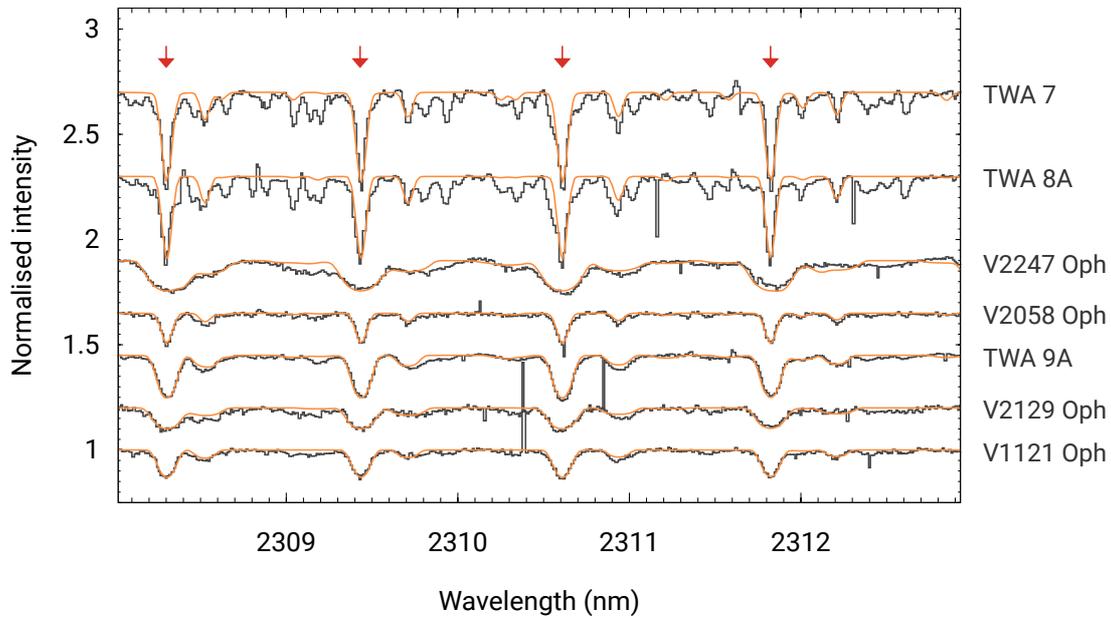}
\caption{
    Normalised CRIRES spectra (black histogram) in the wavelength setting  $24/-1/n$ covering the magnetically insensitive CO lines used to measure {\vsini} (marked with red arrows). The corresponding synthetic spectra computed with the best-fitting {\vsini} values are overplotted in orange.
}
\label{figure:vsini}
\end{figure*}

The stellar parameters that are necessary inputs for our analysis are the effective temperature {\teff}, the surface gravity $\log g$, the projected rotational velocity {\vsini}, and both the micro- and macrotubulent velocities, {\vmic} and {\vmac}. Effective temperatures were adopted from the literature according to the information in Table~\ref{table:parameters}. Surface gravities (rounded to 0.5 dex) were determined from the position of stars on the HR diagram, as illustrated in Fig.~\ref{figure:yapsi-logg}. We then adopted a {\tt MARCS} model atmosphere with the closest {\teff} and log~$g$ values. For stars that we observed in the wavelength setting $24/-1/n$, we determined {\vsini} through Markov chain Monte Carlo (MCMC) fitting of the magnetically insensitive CO lines present in that wavelength region. The free parameters that were allowed to vary were {\vsini} and a scaling factor $k$. The observed spectra and synthetic calculations with the best-fit {\vsini} values are shown in Fig.~\ref{figure:vsini}. The typical errors on {\vsini} introduced by changes in the stellar parameters are: $\pm 0.1$~{\kms} for {\teff} changes of $\pm 200~K$, and $\pm 0.2$~{\kms} for $\log g$ variations of $\pm 0.5$~dex. For TW Hya, we adopted a {\vsini} value from \citet{2011donati-twhya}. We assumed fixed values of $v_{\text{mic}} = 2$~{\kms} and $v_{\text{mac}} = 0$~{\kms} following \citet{2017A&A...608A..77L}. The stellar parameters for all targets are listed in Table~\ref{table:parameters}.

\begin{table*}
\caption{Stellar parameters.}
\label{table:parameters}
\centering
\begin{tabular}{r c l l l}     
\hline\hline
Target ID   & {\teff}                   & log$g$                    & {\vsini}                  & References                    \\
            & (K)                       &                           & (\kms)                    &                \\
\hline
TWA 7       & 3300~\tablefootmark{a}    & 4.0  & $4.1^{+0.1}_{-0.1}$   & \citet{2008AJ....136.2286Y}   \\
TWA 8A      & 3400~\tablefootmark{a}    & 4.0   & $4.4^{+0.2}_{-0.2}$   & \citet{2008AJ....136.2286Y}   \\
V2247 Oph   & 3500~\tablefootmark{a}    & 3.5   & $19.5^{+0.1}_{-0.1}$  & \citet{2010donati-v2247oph}   \\
V2058 Oph   & 3800~\tablefootmark{a}    & 3.5   & $5.5^{+0.1}_{-0.1}$   & \citet{2015AA...579A..66M}    \\
TWA 9A      & 4000~\tablefootmark{a}    & 4.0   & $10.6^{+0.3}_{-0.3}$  & \citet{2008AJ....136.2286Y}   \\
TW Hya      & 4075~\tablefootmark{a}    & 4.0   & 4.0~\tablefootmark{a} & \citet{2011donati-twhya}      \\
V2129 Oph   & 4500~\tablefootmark{a}    & 3.5   & $14.6^{+0.7}_{-0.6}$  & \citet{2007MNRAS.380.1297D}   \\
V1121 Oph   & 4500~\tablefootmark{b}    & 3.5   & $9.7^{+0.1}_{-0.1}$   & \citet{2010AA517A88W}         \\
\hline
\end{tabular}
\tablefoot{
    The {\vsini} values are given with error bars corresponding to the $1\sigma$ confidence interval.
    \tablefoottext{a}{Adopted from the reference in the last column.} 
    \tablefoottext{b}{V1121~Oph has the same reported spectral type (K5) as V2129~Oph. For this reason, we adopt the same {\teff} value than V2129~Oph.}
    
}
\end{table*}

\subsection{Magnetic field analysis}
\label{subsection:magnetic_field_analysis}
To determine the magnetic field modulus {\bb}, we fitted magnetically sensitive lines in the observed spectra with magnetic synthetic spectra. These synthetic spectra were generated using different prescriptions for the distribution of surface magnetic field strength. We started by adopting the same three prescriptions as in \citet{2008AJ....136.2286Y}, also following their model nomenclature (Model 1, 2, and 3).

In Model~1, a uniform magnetic field $B$ covers a fraction of the stellar surface. The rest of the surface is field-free. The free parameters of this model include the magnetic field strength $B$ in the grid from 0 to 14~kG and the magnetic filling factor $f$. The magnetic field modulus is given by {\bb} $ = Bf$. In Model~2, the stellar surface is split into three regions: one non-magnetic region and two magnetic regions with variable field strengths $B_1$ and $B_2$ (within the 0--14 kG grid) and corresponding filling factors $f_1$ and $f_2$. The free parameters of this model are the two magnetic filling factors and the two field strength values. The magnetic field modulus is calculated as {\bb} $= B_1 f_1 + B_2 f_2$. Model~3 postulates three magnetic regions on the stellar surface with fixed magnetic field strengths $B_1, B_2, B_3$ set respectively to 2, 4, and 6 kG. The remainder of the surface is non-magnetic. The free parameters of this model are the three magnetic filling factors $f_1$, $f_2$, $f_3$. The resulting magnetic field modulus is given by {\bb} $= \sum_{i = 1}^{3} B_i f_i$.

Furthemore, we found that, given the high quality of our observations, proper fitting of spectral lines of some of the stars in our sample required fields stronger than 6 kG. Consequently, we introduced a new prescription (Model 4), representing a generalisation of Model 3 to include stronger field components. Model 4 allows up to seven magnetic regions on the stellar surface with field strengths $B_1$, $B_2$, $B_3$, $B_4$, $B_5$, $B_6$, and $B_7$ set to 2, 4, 6, 8, 10, 12, and 14 kG while the rest of the surface is non-magnetic. The number of magnetic components is selected via an information criterion within the framework described in Sect.~\ref{subsection:mcmc}. The free parameters of this model is the set of magnetic filling factors $f_i$; the magnetic field modulus is given by {\bb} $ = \sum_{i = 1}^{7} B_i f_i$.

The spectra of T Tauri stars are known to be affected by veiling, which often significantly reduces the residual line depths. In addition, some of the stars might exhibit deviations from the solar chemical composition assumed in our calculations. It is impossible to disentangle these effects based on the short wavelength intervals observed with CRIRES. For this reason, we introduced a single parameter $k$ (scaling factor) that accounts for both the true veiling and other deviations in residual line depths (due to e.g. chemical composition, continuum placement). The scaling factor $k$ modifies the synthetic spectra $s$ according to the relation $s_{\text{scaled}} = s \cdot k +1 -k$. This parameter was optimised together with the magnetic parameters for each of the four field parameterisation models described above.

In practice, our analysis consisted of: firstly calculating a library of synthetic spectra using a given set of stellar parameters and a grid of magnetic field strengths with the {\tt Synmast} code, secondly convolving these spectra for appropriate non-magnetic broadening ({\vsini}, instrumental broadening corresponding to $R=10^5$) using the {\tt s3div} code\footnote{\url{http://www.astro.uu.se/~oleg/synth3.html}}, thirdly combining the individual magnetic synthetic spectra into a composite synthetic spectrum following one of the four models described above, fourthly fitting the observed spectra by performing a full-grid search (models 1--2) or a MCMC search (models 3--4) in the free-parameter space. The scripts and example data used to perform the analysis are available on GitHub\footnote{\url{https://github.com/astro-alexis/magnotron-tts}}.

\subsection{Models 1--2: full-grid search}
\label{subsection:fullgrid}
We generated a grid of synthetic magnetic field spectra for the set of stellar parameters ({\teff}, {\vsini}, and $\log g$) of our sample, with the magnetic field strength $B$ ranging from 0 to 14~kG in increments of 1~kG.  For each of the two magnetic field prescriptions, we then conducted a search for a set of free parameters which minimised the $\chi_r^2$ of the fit between observed and synthetic spectra. The scaling factor $k$ was allowed to range from $0$ to $2$ in increments of $10^{-2}$ , the magnetic filling factors varied from $0$ to $1$ in increments of $2 \times 10^{-2}$ and $4 \times 10^{-2}$ for Models 1 and 2, respectively. The non-magnetic filling factor was not a free parameter, as it is the remainder of the surface that is not covered with a magnetic field. This translated to an additional constraint that the sum of all filling factors must be equal to unity. 

\subsection{Model 3--4: MCMC optimisation}
\label{subsection:mcmc}
Our MCMC analysis made use of the Goodman \& Weare's affine invariant ensemble (AIE) sampler \citep{2010CAMCS...5...65G} in its implementation in the Python package {\tt emcee} \citep{2013PASP..125..306F}. The free parameters of Models~3--4 were the magnetic filling factors $f_i$ of the magnetic regions and the scaling factor $k$. We adopted a uniform prior $U(0,1)$ for each of the individual magnetic filling factors $f_i$ and an additional prior $U(0,1)$ for the sum $\sum_{i=1}^{N} f_i$. This prior means that the fraction of the stellar surface covered by any magnetic field must be less than or equal to unity, the remainder was set non-magnetic. 
A uniform prior $U(0,2)$ was adopted for the scaling factor. We generated a set of randomised filling-factor combinations as an initial value for each Markov chain. The initial value of the scaling factor was set to $k = 0.8$ with an added random noise. A simple Gaussian likelihood was used to measure the deviation of the model to the observed data adopting a standard deviation $\sigma = \text{S/N}^{-1}$.

The MCMC AIE sampler was run with 60 chains, each initially 1000 steps long. We discarded the first 800 `burn-in' samples after ensuring that each chain had reached a stable state well within the burn-in length. Formally, the MCMC was iteratively run with 1000 additional steps until the chains for each parameter reached a minimum of 2500 effective sample size (ESS) defined as
$$ \text{ESS} =\frac{N}{2\tau_{\text{int}}}  $$
where $N$ is the total number of samples in the chains, and $\tau_{\text{int}}$ is the integrated autocorrelation \citep[][section 3.7.1]{sharma2017}. The resulting jump acceptance fractions were comprised between 44 and 65\% (median: 53\%).

For each observed spectrum, we ran the MCMC analysis as described above multiple times using field strength parameterisation models of increasing complexity. We started with the simplest model comprised of only two components: 0 and 2~kG. We iteratively added more components: 4, 6, 8, 10, 12, and ultimately 14~kG. More complex models generally provide a better fit to observations, but can also overestimate the magnetic field modulus by introducing a small fraction of very strong field that effectively fits the noise or systematics in the spectra. To avoid this issue, we selected the model complexity that is favoured by the data by choosing the model that yields the smallest Bayesian Information Criterion (BIC). The BIC is a model-selection tool that estimates the relative quality of a model applied for fitting a given dataset with regards to the number of model parameters. The criterion is defined as $$BIC \equiv -2 \text{ln}\mathcal{L}_{\text{max}} +2 n\text{ ln}N,$$ where $\mathcal{L}_{\text{max}}$ is the maximum likelihood value, $n$ is the number of model parameters, and $N$ is the number of data points  \citep{sharma2017}.
\begin{figure}
\centering
\includegraphics[width=0.9\columnwidth]{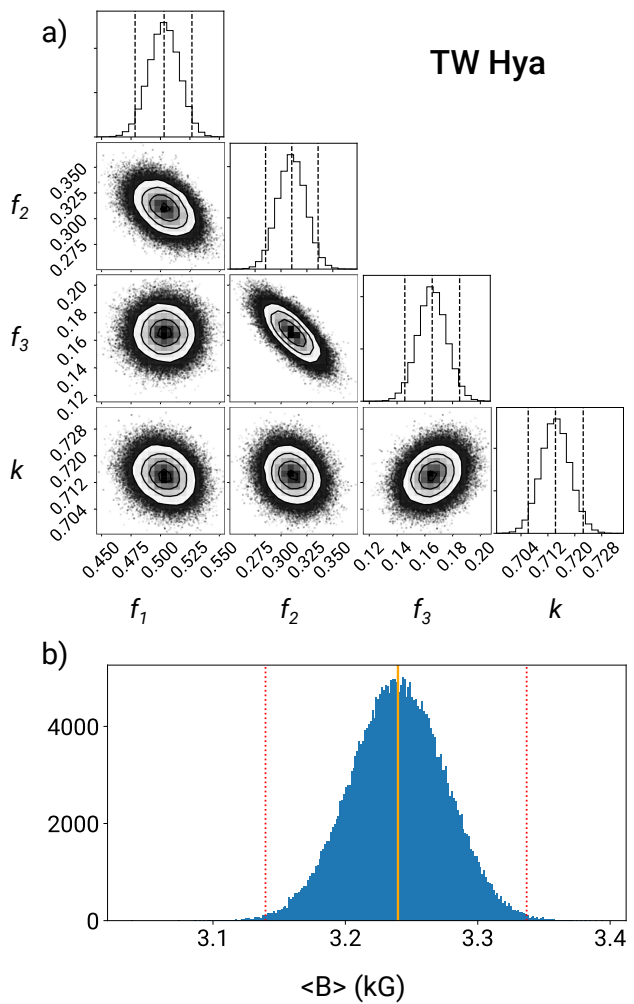}
\caption{
    Illustration of the MCMC analysis for the spectrum of TW~Hya obtained on 2009-02-12 using Model~4. The corner plot at the top shows the posterior distribution of the model parameters (magnetic filling factors $f_i$ and the scaling factor $k$). The bottom histogram displays the resulting posterior distribution of the magnetic field modulus {\bb}, overplotted with its median value (yellow thick line) and the limits of the $3\sigma$ confidence interval (red dashed lines).
}
\label{figure:mcmc}
\end{figure}

\section{Results}
\label{section:results}
We obtained a set of four magnetic field modulus {\bb} values for each observation, corresponding to each of the four magnetic models. These results are presented in Table~\ref{table:b}, which summarises magnetic measurements together with the corresponding reduced $\chi_r^2$ values. The simplest Model~1 fails to reproduce the observed spectra satisfactorily in many cases yielding high $\chi_r^2$ values. For stars with moderate field strengths ($\la$\,3~kG), Models 2 and 3 produce an acceptable fit to the data, with $\chi_r^2$ values very close to those obtained with Model~4. However, these parameterisations lead to larger $\chi_r^2$ for stars with stronger fields. Model~4 appears to provide a reliable fit to all spectra in our sample, up to high magnetic field strengths ($>4$~kG), without overfitting spectra of stars with weak apparent magnetic field (e.g. V2058~Oph). For this reason, we judge Model~4 to be the most reliable and will use results obtained with this model in the remainder of the paper. 

\begin{figure*}
\centering
\includegraphics[width=1.5\columnwidth]{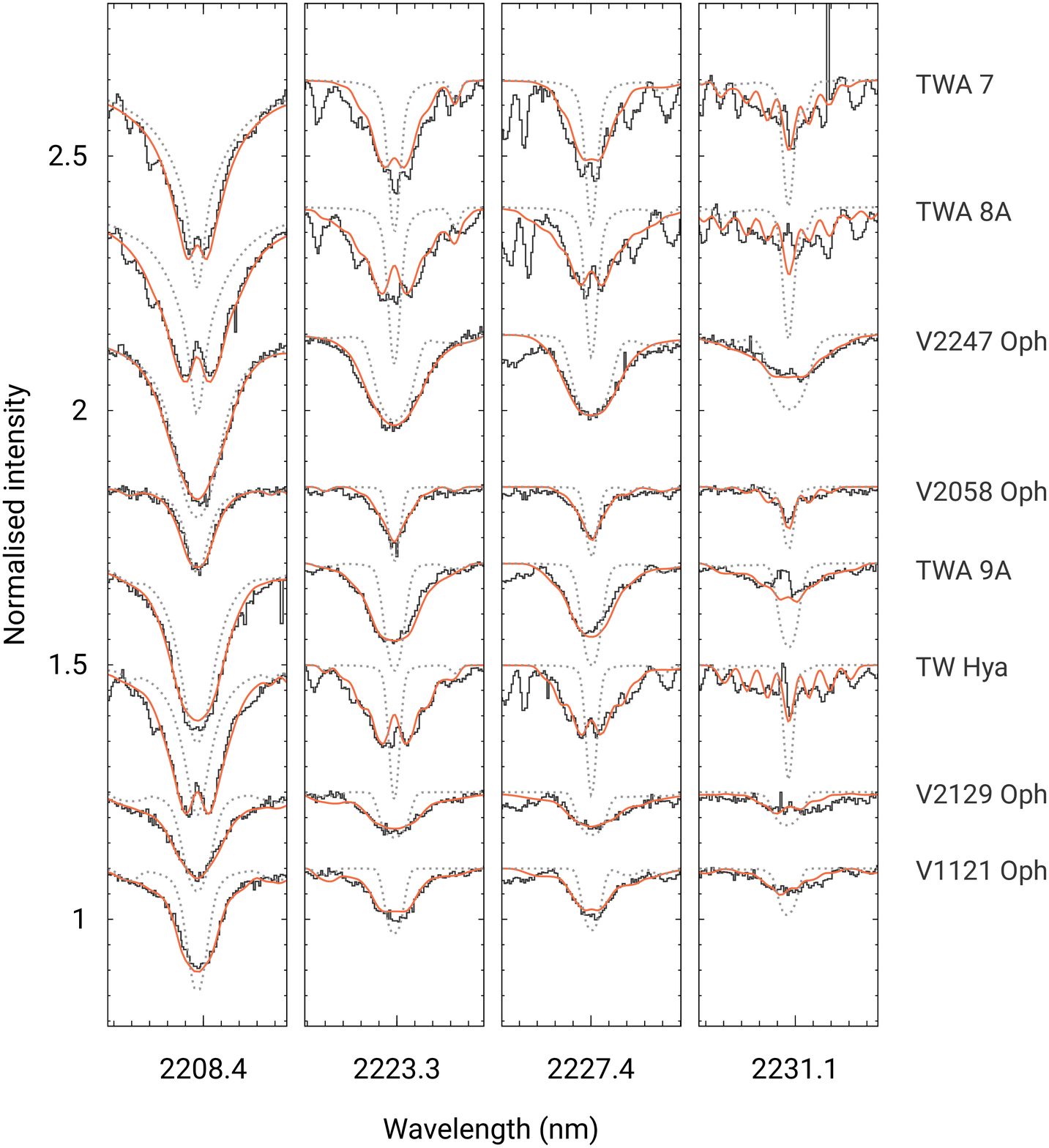}
\caption{
    Highest-S/N observation of each target around the four magnetically-sensitive spectral lines used in this analysis (black histogram), overplotted with the best non-magnetic fit (grey dashed line) and the best magnetic fit with model~4 (orange line).
}
\label{figure:magnetic-fit}
\end{figure*}

\begin{figure}
\centering
\includegraphics[width=1.0\columnwidth]{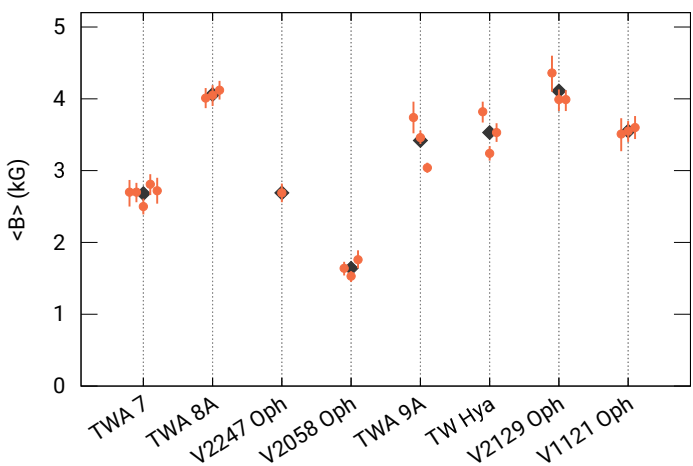}
\caption{
    Individual {\bb} measurements for each target from model~4 (filled orange circles with $3\sigma$ error bars) and average {\bb} value for each star (black diamonds).
}
\label{figure:rot}
\end{figure}

A typical parameter inference result from the application of Model~4 is displayed in Fig.~\ref{figure:mcmc}, where the top subplot shows a corner plot with the posterior probability distributions for the model parameters, and the bottom subplot shows the posterior probability distribution of the surface magnetic field {\bb}. Typical $99.7$\% confidence intervals for {\bb} obtained from the MCMC analysis have a width of $0.3$ kG. However, these statistical uncertainties can be smaller than the systematic uncertainties that would arise from errors of the adopted atomic data, such as the Van der Waals broadening parameters. The typical errors on the magnetic field modulus introduced by errors on stellar parameters are: $\pm 0.02$~kG for {\vsini} variations of $\pm 0.5${\kms}, $\pm 0.1$~kG for {\teff} changes of $\pm 200$~$K$, and $\pm 0.3$~kG for surface gravity errors of $\pm 0.5$~dex.

The best fits obtained with Model~4 to the highest-S/N observations of each target are illustrated in Fig.~\ref{figure:magnetic-fit}. For comparison, this figure also shows non-magnetic synthetic spectra calculated with the same stellar parameters. It is evident that, in all cases, magnetic field has a large impact on the studied spectral lines.

By obtaining repeated observations and magnetic field measurements for all stars (with the exception of V2247~Oph), we could assess the degree of rotational variability of the magnetic field modulus {\bb}. This analysis establishes a relatively low level of night-to-night changes of the mean field strength. Fig.~\ref{figure:rot} displays individual {\bb} measurements and the average value for each star. The maximum deviation of an individual measurement from the average is 0.4~kG (TWA~9A). The mean deviation is 0.1~kG. The maximum peak-to-peak variation of individual {\bb} measurements for a given star is 0.7~kG (for TWA~9A, which amounts to a $21\%$ variation relative to the mean value of {\bb}\,$=3.4$~kG). The mean peak-to-peak variation is 0.3~kG. The absence of dramatic magnetic field changes with the stellar rotation indicates that magnetic regions responsible for the Zeeman broadening signal are not concentrated in discrete patches but distributed approximately homogeneously over the stellar surface.

\begin{table*}
\caption{
    Magnetic field modulus {\bb} determined through our analysis using the four different magnetic field prescriptions.
}
\label{table:b}
\centering
\begin{tabular}{r c c | c | c | c c c c}     
\hline\hline
            & & Model 1           & Model 2           & Model 3           & Model 4           & & &                    \\
Target ID   & UT date& {\bb} ($\chi_r^2$)  & {\bb} ($\chi_r^2$)  & {\bb} ($\chi_r^2$)  & {\bb} ($\chi_r^2$)  & {\bb}$_\text{min}$    & {\bb}$_\text{max}$    & $B_{\text{max}}$   \\
            & & (kG)              &(kG)               & (kG)              & (kG)              & (kG) & (kG) & (kG)           \\
\hline
TWA 7     	&2008-12-24	& 2.3 (3.8) 	& 2.4 (3.5)    	& 2.7 (3.6) 	& 2.7 (3.5) 	& 2.5 	& 2.9 	& 4 \\ 
TWA 7     	&2008-12-24	& 2.3 (5.9) 	& 2.7 (5.1)    	& 2.7 (5.0) 	& 2.7 (5.0) 	& 2.6 	& 2.8 	& 4 \\ 
TWA 7     	&2009-02-12	& 2.2 (6.4) 	& 2.5 (4.9)    	& 2.6 (4.9) 	& 2.5 (4.9) 	& 2.4 	& 2.6 	& 4 \\ 
TWA 7     	&2008-12-24	& 2.4 (7.2) 	& 2.8 (6.1)    	& 2.8 (6.1) 	& 2.8 (6.0) 	& 2.7 	& 2.9 	& 4 \\ 
TWA 7     	&2009-02-24	& 2.0 (12.0) 	& 2.6 (9.8)    	& 2.7 (9.7) 	& 2.7 (9.7) 	& 2.5 	& 2.9 	& 6 \\ 
TWA 8A    	&2009-02-26	& 3.1 (7.3) 	& 3.6 (4.6)    	& 3.7 (3.7) 	& 4.0 (3.4) 	& 3.9 	& 4.2 	& 8 \\ 
TWA 8A    	&2009-02-27	& 3.2 (7.4) 	& 3.8 (4.7)    	& 3.7 (3.8) 	& 4.0 (3.5) 	& 3.9 	& 4.2 	& 8 \\ 
TWA 8A    	&2009-02-28	& 3.1 (7.8) 	& 3.7 (4.7)    	& 3.7 (3.7) 	& 4.1 (3.3) 	& 4.0 	& 4.3 	& 8 \\ 
V2247 Oph 	&2008-08-24	& 2.5 (3.2) 	& 2.8 (2.6)    	& 2.7 (2.5) 	& 2.7 (2.5) 	& 2.6 	& 2.8 	& 6 \\ 
V2058 Oph 	&2008-07-09	& 1.3 (3.5) 	& 1.8 (2.3)    	& 1.7 (2.4) 	& 1.6 (2.4) 	& 1.5 	& 1.7 	& 4 \\ 
V2058 Oph 	&2008-07-10	& 1.2 (4.9) 	& 1.5 (3.2)    	& 1.5 (3.2) 	& 1.5 (3.2) 	& 1.5 	& 1.6 	& 4 \\ 
V2058 Oph 	&2008-08-26	& 1.3 (4.4) 	& 1.7 (2.5)    	& 1.8 (2.5) 	& 1.8 (2.5) 	& 1.6 	& 1.9 	& 6 \\ 
TWA 9A    	&2008-07-08	& 2.6 (7.2) 	& 3.4 (5.3)    	& 3.3 (5.3) 	& 3.7 (5.0) 	& 3.5 	& 4.0 	& 10 \\ 
TWA 9A    	&2009-02-27	& 2.6 (16.5) 	& 3.1 (7.0)    	& 3.2 (5.8) 	& 3.5 (5.5) 	& 3.4 	& 3.6 	& 10 \\ 
TWA 9A    	&2009-02-28	& 2.2 (19.4) 	& 2.8 (11.1)    	& 2.9 (9.8) 	& 3.0 (9.5) 	& 3.0 	& 3.1 	& 8 \\ 
TW Hya    	&2008-12-24	& 2.6 (10.6) 	& 3.6 (5.8)    	& 3.6 (4.6) 	& 3.8 (4.5) 	& 3.7 	& 4.0 	& 8 \\ 
TW Hya    	&2009-02-12	& 2.5 (10.7) 	& 3.2 (7.1)    	& 3.2 (6.4) 	& 3.2 (6.4) 	& 3.1 	& 3.3 	& 6 \\ 
TW Hya    	&2009-02-15	& 2.6 (10.8) 	& 3.5 (5.6)    	& 3.4 (4.0) 	& 3.5 (4.0) 	& 3.4 	& 3.7 	& 8 \\ 
V2129 Oph 	&2008-08-24	& 2.6 (5.2) 	& 4.1 (2.8)    	& 3.4 (3.5) 	& 4.4 (2.6) 	& 4.1 	& 4.6 	& 12 \\ 
V2129 Oph 	&2008-08-26	& 2.8 (5.6) 	& 3.6 (4.1)    	& 3.6 (4.2) 	& 4.0 (4.0) 	& 3.8 	& 4.1 	& 8 \\ 
V2129 Oph 	&2008-08-28	& 2.6 (6.1) 	& 3.9 (4.4)    	& 3.6 (4.7) 	& 4.0 (4.3) 	& 3.8 	& 4.1 	& 8 \\ 
V1121 Oph 	&2008-08-23	& 2.0 (5.8) 	& 3.0 (3.3)    	& 2.9 (3.3) 	& 3.5 (2.9) 	& 3.3 	& 3.7 	& 10 \\ 
V1121 Oph 	&2008-08-25	& 2.1 (10.0) 	& 3.0 (4.9)    	& 3.0 (4.6) 	& 3.6 (3.4) 	& 3.4 	& 3.7 	& 10 \\ 
V1121 Oph 	&2008-07-24	& 2.1 (8.7) 	& 3.4 (3.6)    	& 3.0 (4.1) 	& 3.6 (2.8) 	& 3.4 	& 3.8 	& 10 \\ 

\hline
\end{tabular}
\tablefoot{The $\chi^2_r$ value associated with each fit is indicated within parentheses. For model~4, the extrema of the 3$\sigma$ confidence intervals {\bb}$_\text{min}$ and {\bb}$_\text{max}$ are also indicated, as well as the maximum magnetic field value $B_{\text{max}}$ allowed in the model selected through the minimisation of the Bayesian Information Criterion.}
\end{table*}

\section{Discussion}
\label{section:discussion}
\begin{figure*}
\centering
\includegraphics[width=2.0\columnwidth]{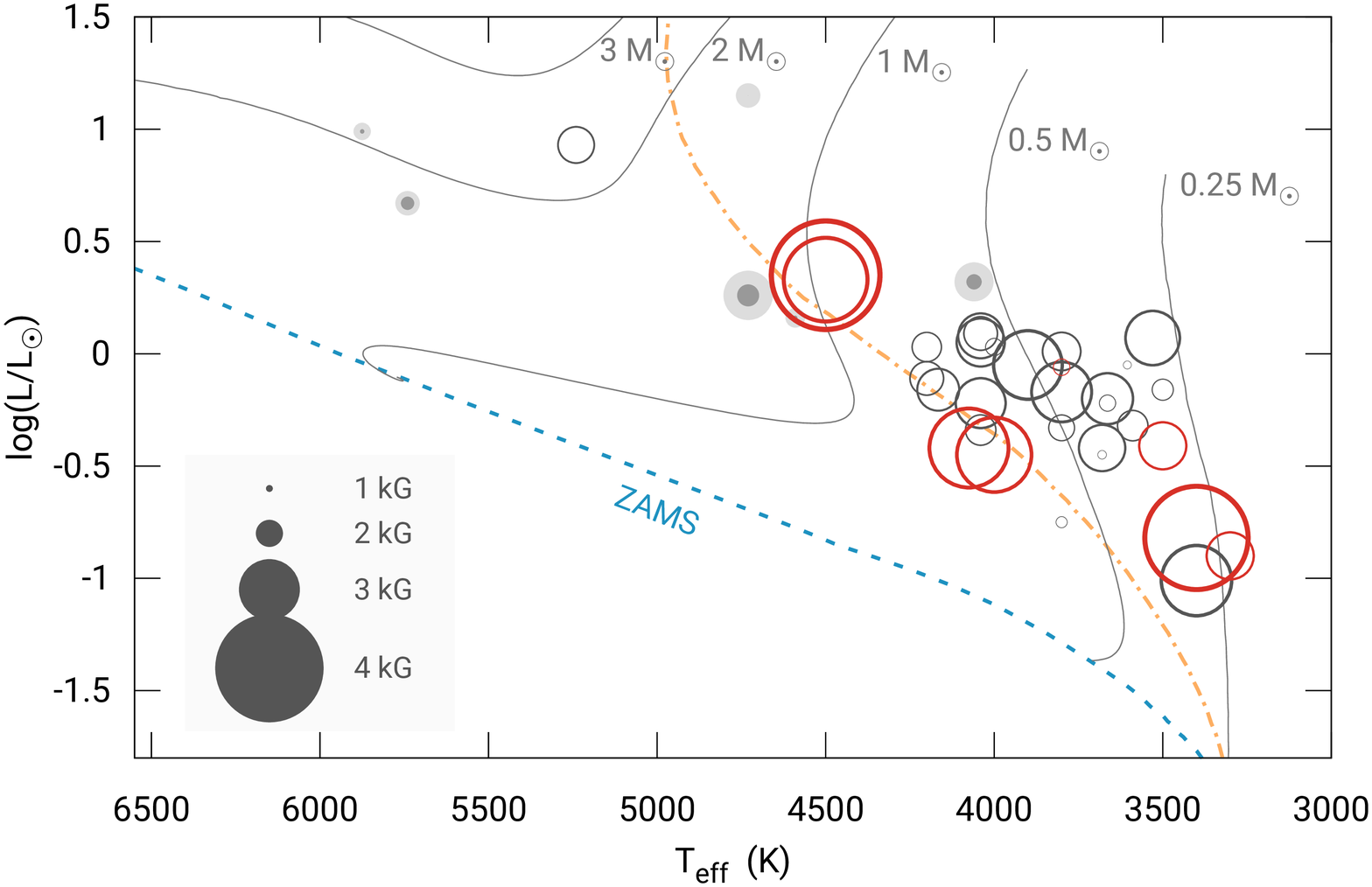}
\caption{
    HR diagram summarizing near-infrared spectroscopic magnetic field measurements of T Tauri stars. The filled grey circles represent the intermediate-mass T Tauri stars sample from \citet{2017A&A...608A..77L}. The two shades of grey represent results obtained for two different sets of spectral lines. The open dark circles show the samples from \citet{2007ApJ...664..975J}, \citet{2008AJ....136.2286Y}, and \citet{2011ApJ...729...83Y}, and the open red circles display the average measurement for the sample presented in this work. The radius of the circles is proportional to $\langle B \rangle^2$.  The Zero Age Main-Sequence (ZAMS) is plotted with dashed blue line, the limit between fully and partially convective regimes is shown with the dash-dotted yellow curve, and evolutionary tracks for different masses are plotted with thin grey lines.
}
\label{figure:yapsi}
\end{figure*}

In order to put our results in a wider context, we present our mean magnetic field measurement for each star in the sample on the Hertzsprung-Russell (H-R) diagram alongside literature measurements of Zeeman broadening from near-infrared spectra of other low mass and intermediate mass T Tauri stars, as shown in Fig.~\ref{figure:yapsi}. The mass tracks, the zero-age main sequence (ZAMS), and the fully-convective transition boundary displayed in the H-R diagram are all extracted from the Yale–Potsdam Stellar Isochrones grid \citep{2017ApJ...838..161S}, which is available online\footnote{\url{http://www.astro.yale.edu/yapsi/}}. It is worthwile to note that this stellar evolution model does not include the effect of stellar magnetic fields, which are making a PMS star appear cooler and therefore potentially younger. Incorporating magnetic fields into stellar evolution models is an ongoing effort in order to improve their accuracy \citep[e.g. ][]{2016A&A...593A..99F}. To place the stars on the H-R diagram, we used effective temperatures from the literature, and derived consistent stellar luminosities following a methodology adapted from \citet{2019villebrun}. We obtained $J$-band magnitudes $m_J$ from 2MASS \citep{2003tmc..book.....C}, $V$-band magnitudes $m_V$ from \citet{1992A&AS...92..481B} and \citet[and references therein]{2006A&A...460..695T}, and parallaxes $\pi$ from GAIA DR2 \citep{2016A&A...595A...1G,2018A&A...616A...1G}. Then, we obtained the stellar luminosities through the following set of equations:
$$ E(V-J) = (m_V - m_J) - (V-J)_0, $$
$$ A_J =  R_J \times E(V-J), $$
$$ M_J = m_J + 5 + 5 \text{ log}_{10}(\pi) - A_J, $$ 
$$ M_\text{bol} = M_j + BC_J, $$
$$ \frac{L}{L_\sun} = 10^{-(M_{\text{bol}} - M_{\text{bol,\sun}})/2.5}, $$
where $(V-J)_0$ is the empirical intrinsic colour and $E(V-J)$ the colour excess, $A_J$ the extinction in the $J$-band, $R_J$ the total to selective extinction set to 0.437 following \citet{2019villebrun}, $\pi$ the parallax in arcseconds, and $BC_J$ the bolometric correction in the $J$-band. The empirical intrinsic colours and bolometric corrections for 5--30~Myr PMS stars were computed according to \citet{2013ApJS..208....9P}, and we adopted the associated value of $M_{\text{bol,\sun}} = 4.755$. The determined luminosities and intermediate values from the computation are listed in Table~\ref{table:luminosities}. It is important to note that the stellar luminosities of these objects can be quite uncertain, due to e.g. their variability, extinction, and the non-contemporaneous V- and J-band measurements.

From examination of Fig.~\ref{figure:yapsi} it appears that the magnetic field modulus is not a function of the position in the H-R diagram, at least in the fully convective regime where we have a substantial number of observations showing stars in the same location having quite different magnetic field modulus {\bb}. Across the fully convective boundary the sample of stars is small. The two stars from our sample in this regime (albeit quite close to the boundary) are TWA~9A and TW~Hya with mean field modulus of \bb $= 3.4$ and \bb $= 3.5$~kG respectively. This is stronger compared to the field modulus reported for the intermediate mass T Tauri stars sample of \citet{2017A&A...608A..77L}, in which four stars are partially convective.

\begin{table*}
\caption{Stellar luminosities.}
\label{table:luminosities}
\centering
\begin{tabular}{r c c c c c c c c c c}     
\hline\hline
Target ID   & GAIA DR2 ID           & Parallax  & $m_J$     & $m_V$                     & $(BC)_J$  & $(V-J)_0$ & $E(V-J)$                  & $A_J$                     & $M_{\text{bol}}$  & $L$           \\
            &                       & (mas)     & (mag)     & (mag)                     & (mag)     & (mag)     & (mag)                     & (mag)                     & (mag)             & ($L_\sun)$    \\
\hline
TWA 7       & 5444751795151480320   & 29.4      & 7.79      & 11.70~\tablefootmark{a}   & 1.86      & 4.00      & $0.00$~\tablefootmark{c}  & $0.00$~\tablefootmark{c}  & $6.99$            & $0.13$        \\ 
TWA 8A		& 3485098646237003392   & 21.6      & 8.34      & 12.20~\tablefootmark{a}   & 1.83      & 3.78      & $0.08$                    & $0.03$                    & $6.81$            & $0.15$        \\ 
V2247 Oph   & 6049072007438134272   & 8.9       & 9.42      & 13.40~\tablefootmark{b}   & 1.80      & 3.58      & $0.40$                    & $0.17$                    & $5.79$            & $0.39$        \\ 
V2058 Oph   & 6049142410542091648   & 7.4       & 9.15      & 12.80~\tablefootmark{b}   & 1.68      & 3.04      & $0.61$                    & $0.27$                    & $4.91$            & $0.87$        \\ 
TWA 9A      & 3463395519357786752   & 13.1      & 8.68      & 11.30~\tablefootmark{a}   & 1.62      & 2.73      & $0.00$~\tablefootmark{c}  & $0.00$~\tablefootmark{c}  & $5.88$            & $0.35$        \\ 
TW Hya      & 5401795662560500352   & 16.6      & 8.22      & 11.10~\tablefootmark{a}   & 1.60      & 2.62      & $0.26$                    & $0.11$                    & $5.81$            & $0.38$        \\
V2129 Oph   & 6049153921054413440   & 7.7       & 8.44      & 11.60~\tablefootmark{a}   & 1.46      & 2.11      & $1.05$                    & $0.46$                    & $3.87$            & $2.26$        \\
V1121 Oph   & 4326521359869384576   & 8.3       & 8.30      & 11.40~\tablefootmark{a}   & 1.46      & 2.11      & $0.99$                    & $0.43$                    & $3.92$            & $2.16$        \\
\hline
\end{tabular}
\tablefoot{
    \tablefoottext{a}{\citet[and references therein]{2006A&A...460..695T}}
    \tablefoottext{b}{\citet{1992A&AS...92..481B}}
    \tablefoottext{c}{These colour excesses and extinctions were set to $0$. The original computed values of $E(V-J)$ for TWA~7 and TWA~9A were respectively $-0.09$ and $-0.11$.}
}
\end{table*}

We can also compare the magnetic field modulus {\bb} from this work, with the average large-scale magnetic field $\langle B_V \rangle$ obtained from Stokes $V$ ZDI studies. Values for $\langle B_V \rangle$ are $113$~G for TWA~9A \citep{2018MNRAS.480.1754N}, 1628~G for TW~Hya \citep{2011donati-twhya}, $\approx$\,300~G for V2247~Oph \citep{2010donati-v2247oph}, $1424$~G for TWA~8A \citep{2019MNRAS.tmp..410H}, $56$~G for TWA~7 (Nicholson et al. in prep.), and we computed $\langle B_V \rangle \approx 1020$~G from the magnetic field configuration described in \citep{2011donati-v2129oph} for V2129~Oph.

On the one hand, a relatively large fraction (around, respectively, 42\% and 25\% of the mean field strength) of the fields of TW Hya and V2129~Oph are recovered from Stokes $V$. The ZDI mapping indicates that these fields are relatively strong and simple, being mostly poloidal, axisymmetric, and dominated by octupole and dipole components \citep{2011donati-twhya,2011donati-v2129oph}. A large fraction (34\%) of the field of TWA~8A is also recovered from Stokes $V$, with the ZDI mapping also indicating a mostly poloidal field, with most of the poloidal component being axisymmetric \citep{2019MNRAS.tmp..410H}. These values are significantly higher than  those reported for M dwarfs: less than $25\%$ and typically in the $5$--$20$\% range \citep{2009A&A...496..787R,2012LRSP....9....1R,2019ApJ...873...69K}. On the other hand, a smaller fraction of the field is recovered by the Stokes $V$ analyses for V2247~Oph (11\%), whose field is weaker, more complex, and is comprised mostly of a toroidal and a poloidal non-axisymmetric components \citep{2010donati-v2247oph}, for TWA~7 (2\%) which exhibits a predominantly toroidal and non-axisymmetric field (Nicholson et al. in prep.), and for TWA~9A (3\%), which also exhibits a weak and mostly non-axisymmetric large-scale field, but with a dominant poloidal component \citep{2018MNRAS.480.1754N}. 
\begin{figure}
\centering
\includegraphics[width=1.0\columnwidth]{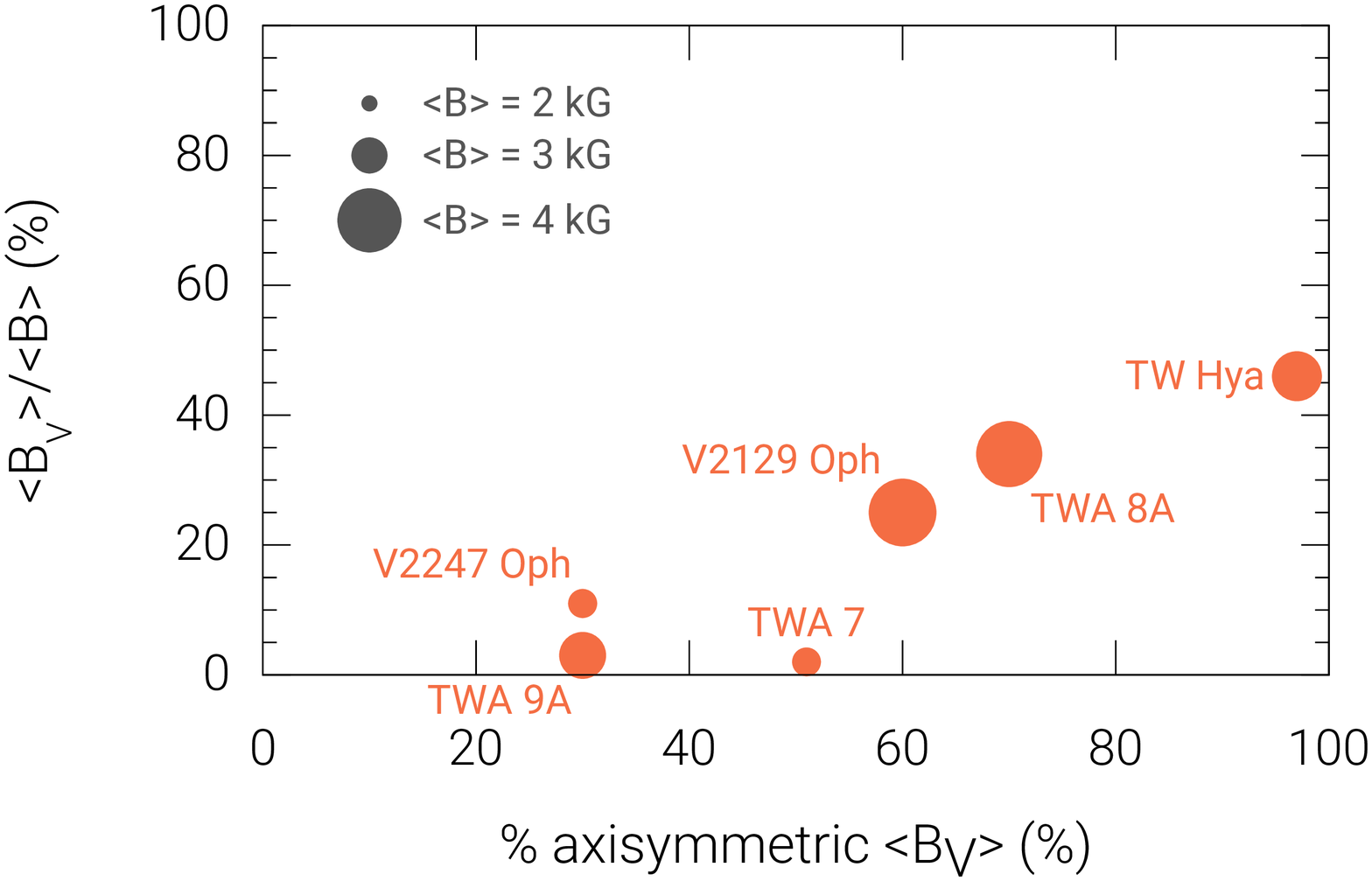}
\caption{
    Fraction of the mean magnetic field recovered by ZDI $\langle B_V \rangle$ over {\bb} as a function of the axisymmetric large-scale magnetic field component recovered with ZDI (magnetic field components in spherical harmonic modes with $m<l/2$). The radius of the symbols is proportional to {\bb}$^2$.}
\label{figure:axisymmetry}
\end{figure}
This assessment seems to corroborate the hypothesis that simple strong axisymmetric fields are better recovered by Stokes $V$ ZDI mapping, but complex fields with structures at smaller scales tend to cancel out and are missed by ZDI, at least when only Stokes $IV$ parameters are used. This is illustrated in Fig.~\ref{figure:axisymmetry}. Naturally, this comparison is somewhat limited by the fact that spectropolarimetric observations used for the tomographic mapping, and spectroscopic observations for this work have been acquired with different instruments and at different epochs. A similar exercise was recently performed by \citet{2019ApJ...876..118S} who collected {\bb} (noted $\langle B_I \rangle$ in their publication) and $\langle B_V \rangle$ values for 83 F-G-K-M stars from multiple literature sources, and derived empirical formulae linking {\bb} and $\langle B_V \rangle$ through power laws. Our {\bb} measurements do not seem to follow the trends with $\langle B_V \rangle$ predicted by either of the two empirical power laws obtained by \citet{2019ApJ...876..118S} (which generally predict too strong fields with {\bb} values as strong as 10 or 14 kG for our sample). Although the discrepancy could possibly arise from the variety of methods used to determine {\bb} values in the literature and different constraints in the ZDI mapping used to recover $\langle B_V \rangle$, it is not at all guaranteed that T Tauri stars should follow the same trend as established using main sequence stars. 

It will be worthwhile to repeat simultaneous determination of {\bb} and $\langle B_V \rangle$ with data from upcoming instruments such as SPIRou or CRIRES+ which will allow one to use both magnetic analysis methods simultaneously on the same dataset. The large spectral grasp of these instruments will also provide the opportunity to analyse molecular features that predominantly form at the cooler spot temperatures so the same dataset can be used to measure both the spot filling factors \citep{2017ApJ...836..200G} and the magnetic filling factors, another clue towards a better understanding of how the extremely strong field is concentrated at the smallest scales. This will enable us to obtain a complete picture of young-star magnetism reconciling our understanding of both small-scale and large-scale magnetic components.

\begin{acknowledgements}
    A.L. would like to thank Colin Hill and Colin Folsom for partaking in useful discussions, and Martin Sahl\'en for helping with the MCMC analysis. A.L is also grateful to Peter K.G. Williams and the contributors to the `Recommendations for Reporting MCMC Analyses in Academic Literature'\footnote{\url{https://github.com/pkgw/mcmc-reporting}} which provided useful guidelines.\\ 
    O.K. acknowledges support by the Knut and Alice Wallenberg Foundation (project grant `The New Milky Way'), the Swedish Research Council (project 621-2014-5720), and the Swedish National Space Board (projects 185/14, 137/17).\\
    Finally, the authors would like to thank the anonymous referee for their report which improved the quality of this work.\\
    Based on observations collected at the European Southern Observatory under ESO programme 081.C-0555.\\
    This work has made use of data from the European Space Agency (ESA) mission {\it Gaia} (\url{https://www.cosmos.esa.int/gaia}), processed by the {\it Gaia} Data Processing and Analysis Consortium (DPAC, \url{https://www.cosmos.esa.int/web/gaia/dpac/consortium}). Funding for the DPAC has been provided by national institutions, in particular the institutions participating in the {\it Gaia} Multilateral Agreement.
    This research has made use of NASA’s Astrophysics Data System Bibliographic Services. \\
    This publication makes use of data products from the Two Micron All Sky Survey, which is a joint project of the University of Massachusetts and the Infrared Processing and Analysis Center/California Institute of Technology, funded by the National Aeronautics and Space Administration and the National Science Foundation.
\end{acknowledgements}
\bibliographystyle{aa}
\bibliography{tts}
\Online

\begin{appendix}
\section{Example of telluric removal with {\tt MOLECFIT}}
\begin{figure*}
\centering
\includegraphics[width=1.9\columnwidth]{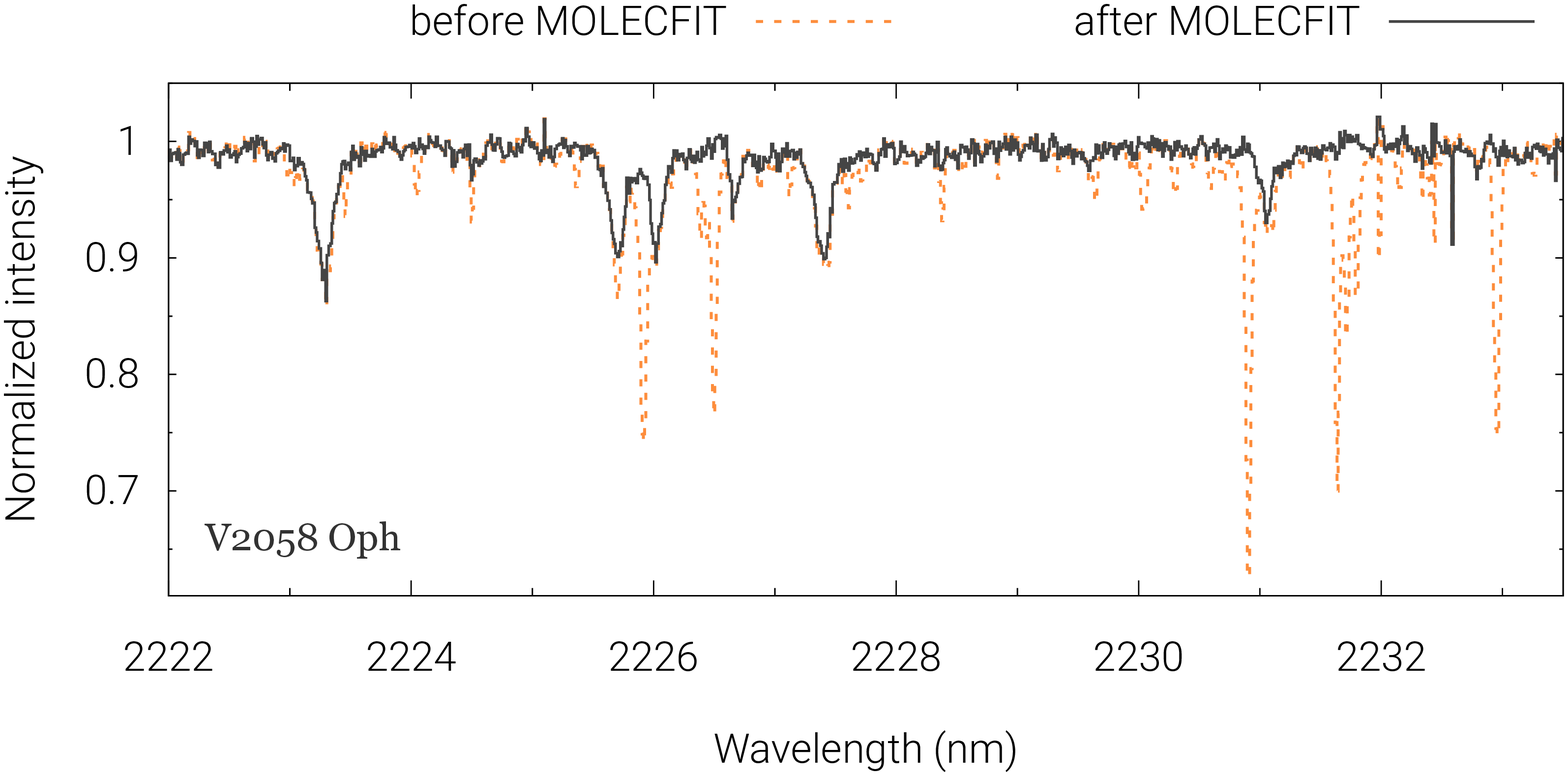}
\caption{
    Example of telluric line removal performed with the {\tt MOLECFIT} software on the spectrum of V2058~Oph acquired on 2008-07-10.
}
\label{figure:molecfit}
\end{figure*}

\section{HR diagram with surface gravity tracks}
\begin{figure*}
\centering
\includegraphics[width=2.0\columnwidth]{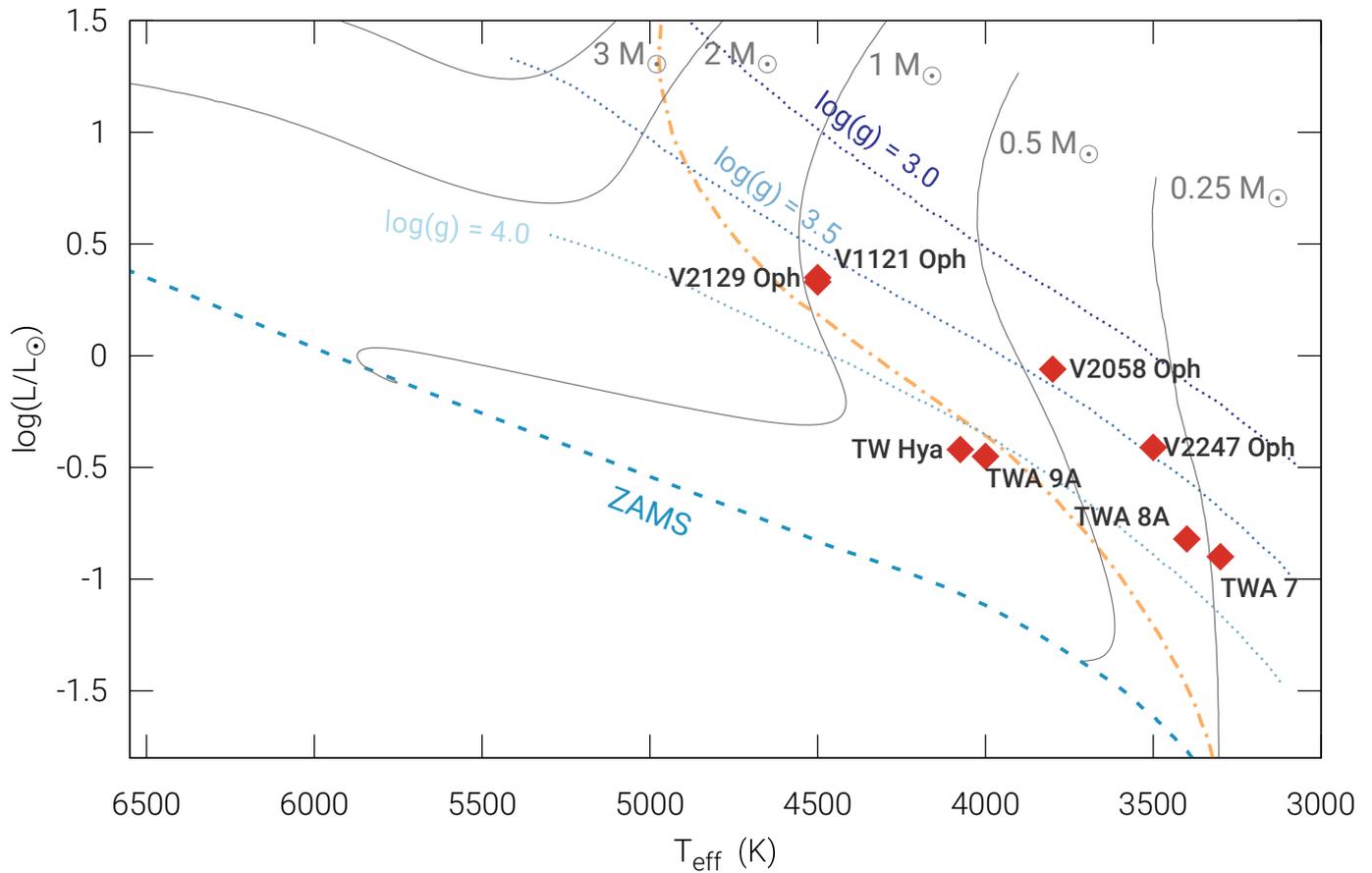}
\caption{
    HR diagram used for surface gravity determination. The red diamonds represent the sample of T Tauri stars presented in this paper. The Zero Age Main-Sequence (ZAMS) is plotted with dashed blue line, the limit between fully and partially convective regimes is shown with the dash-dotted yellow curve. Surface gravity tracks are traced with dotted lines in shades of blue.
}
\label{figure:yapsi-logg}
\end{figure*}

\section{Parameter inference with Model 4}
\begin{table*}
\caption{
    Median parameters obtained from parameter inference with Model~4.
}
\label{table:parameter-inference}
\centering
\begin{tabular}{r c c c  c c c c c c c}     
\hline\hline
Target ID   & UT date & $f_\text{0 kG}$  & $f_\text{2 kG}$  & $f_\text{4 kG}$  & $f_\text{6 kG}$  & $f_\text{8 kG}$  & $f_\text{10 kG}$  & $f_\text{12 kG}$  & $f_\text{14 kG}$  & $k$  \\
            & & & & & & & & & &    \\
\hline
TWA 7       &	2008-12-24 &	0.04	 & 0.57	& 0.39	& -	& -	& -	& -	& -	& 0.73	 \\
TWA 7       &	2008-12-24 &	0.04	 & 0.57	& 0.39	& -	& -	& -	& -	& -	& 0.78	 \\
TWA 7       &	2009-02-12 &	0.10	 & 0.54	& 0.35	& -	& -	& -	& -	& -	& 0.86	 \\
TWA 7       &	2008-12-24 &	0.05	 & 0.50	& 0.45	& -	& -	& -	& -	& -	& 0.78	 \\
TWA 7       &	2009-02-24 &	0.12	 & 0.54	& 0.21	& 0.14	& -	& -	& -	& -	& 0.89	 \\
TWA 8A      &	2009-02-26 &	0.01	 & 0.39	& 0.34	& 0.13	& 0.14	& -	& -	& -	& 0.94	 \\
TWA 8A      &	2009-02-27 &	0.00	 & 0.39	& 0.34	& 0.13	& 0.14	& -	& -	& -	& 0.93	 \\
TWA 8A      &	2009-02-28 &	0.01	 & 0.38	& 0.34	& 0.11	& 0.17	& -	& -	& -	& 0.96	 \\
V2247 Oph   &	2008-08-24 &	0.11	 & 0.50	& 0.33	& 0.06	& -	& -	& -	& -	& 0.79	 \\
V2058 Oph   &	2008-07-09 &	0.35	 & 0.48	& 0.17	& -	& -	& -	& -	& -	& 0.37	 \\
V2058 Oph   &	2008-07-10 &	0.40	 & 0.44	& 0.16	& -	& -	& -	& -	& -	& 0.37	 \\
V2058 Oph   &	2008-08-26 &	0.35	 & 0.45	& 0.16	& 0.04	& -	& -	& -	& -	& 0.37	 \\
TWA 9A      &	2008-07-08 &	0.08	 & 0.40	& 0.28	& 0.10	& 0.06	& 0.07	& -	& -	& 0.89	 \\
TWA 9A      &	2009-02-27 &	0.05	 & 0.48	& 0.27	& 0.14	& 0.02	& 0.04	& -	& -	& 0.84	 \\
TWA 9A      &	2009-02-28 &	0.10	 & 0.51	& 0.24	& 0.09	& 0.07	& -	& -	& -	& 0.82	 \\
TW Hya      &	2008-12-24 &	0.01	 & 0.40	& 0.35	& 0.15	& 0.09	& -	& -	& -	& 0.78	 \\
TW Hya      &	2009-02-12 &	0.02	 & 0.50	& 0.31	& 0.17	& -	& -	& -	& -	& 0.71	 \\
TW Hya      &	2009-02-15 &	0.02	 & 0.43	& 0.35	& 0.15	& 0.05	& -	& -	& -	& 0.77	 \\
V2129 Oph   &	2008-08-24 &	0.03	 & 0.50	& 0.15	& 0.06	& 0.15	& 0.04	& 0.07	& -	& 0.80	 \\
V2129 Oph   &	2008-08-26 &	0.01	 & 0.50	& 0.18	& 0.12	& 0.20	& -	& -	& -	& 0.77	 \\
V2129 Oph   &	2008-08-28 &	0.01	 & 0.49	& 0.23	& 0.03	& 0.24	& -	& -	& -	& 0.81	 \\
V1121 Oph   &	2008-08-23 &	0.14	 & 0.50	& 0.09	& 0.13	& 0.03	& 0.11	& -	& -	& 0.87	 \\
V1121 Oph   &	2008-08-25 &	0.12	 & 0.48	& 0.16	& 0.05	& 0.10	& 0.08	& -	& -	& 0.86	 \\
V1121 Oph   &	2008-07-24 &	0.14	 & 0.44	& 0.18	& 0.02	& 0.14	& 0.07	& -	& -	& 0.87	 \\
\hline
\end{tabular}
\tablefoot{The filling factors are rounded here, but the sum of filling factors should be $\sum_i f_i = 1$.}
\end{table*}
\end{appendix}

\end{document}